\documentclass[aps,prl,twocolumn,superscriptaddress,showpacs,nobibnotes,linenumbers ]{revtex4-2}
\usepackage{graphicx}
\usepackage{amsmath}
\usepackage{amssymb}
\usepackage{epstopdf}
\usepackage{pstool}
\usepackage{float}
\usepackage{pstricks}           
\usepackage{dcolumn}            
\usepackage{bm}                 
\usepackage{hyperref}           
\usepackage{hypernat}           
\usepackage[latin1]{inputenc}   
\usepackage[T1]{fontenc}
\usepackage[english]{babel}
\usepackage{soul}
\usepackage{rotating}
\usepackage{subfigure}
\usepackage{slashed}
\usepackage{color} 

\newcommand*{\re}[1]{\text{Eq. (}\ref{#1}\text{)}}

\newcommand*{\rf}[1]{\text{Fig. }\ref{#1}}

\newcommand*{\frf}[1]{\text{Figure }\ref{#1}}

\begin{document}

\title{Strong-field control of plasmonic properties in core-shell nanoparticles}

\author{Jeffrey Powell}
\thanks{These authors contributed equally to this work.}
\affiliation{J. R. Macdonald Laboratory, Department of Physics, Kansas State University, Manhattan, Kansas 66506, USA}
\affiliation{Department of Physics, University of Connecticut, Storrs, Connecticut 06269, USA}
\affiliation{INRS, {\'E}nergie, Mat{\'e}riaux et T{\'e}l{\'e}communications, 1650 Bld. Lionel Boulet, Varennes, Qu{\'e}bec, J3X1S2, Canada}

\author{Jianxiong Li}
\thanks{These authors contributed equally to this work.}
\affiliation{J. R. Macdonald Laboratory, Department of Physics, Kansas State University, Manhattan, Kansas 66506, USA}
\affiliation{Department of Physics and Astronomy, Louisiana State University, Baton Rouge, Louisiana 70803, USA }

\author{Adam Summers}
\affiliation{J. R. Macdonald Laboratory, Department of Physics, Kansas State University, Manhattan, Kansas 66506, USA}
\affiliation{ICFO - Institut de Ciencies Fotoniques, The Barcelona Institute of Science and Technology, 08860 Castelldefels (Barcelona), Spain}

\author{Seyyed Javad Robatjazi}
\affiliation{J. R. Macdonald Laboratory, Department of Physics, Kansas State University, Manhattan, Kansas 66506, USA}

\author{Michael Davino}
\affiliation{Department of Physics, University of Connecticut, Storrs, Connecticut 06269, USA}

\author{Philipp Rupp}
\affiliation{Physics Department, Ludwig-Maximilians-Universit\"at Munich, D-85748 Garching, Germany}

\author{Erfan Saydanzad}
\affiliation{J. R. Macdonald Laboratory, Department of Physics, Kansas State University, Manhattan, Kansas 66506, USA}

\author{Christopher M. Sorensen}
\affiliation{Department of Physics, Kansas State University, Manhattan, Kansas 66506, USA }

\author{Daniel Rolles}
\affiliation{J. R. Macdonald Laboratory, Department of Physics, Kansas State University, Manhattan, Kansas 66506, USA}

\author{Matthias F. Kling}
\affiliation{Physics Department, Ludwig-Maximilians-Universit\"at Munich, D-85748 Garching, Germany}
\affiliation{Max Planck Institute of Quantum Optics, D-85748 Garching, Germany}

\author{Carlos Trallero-Herrero}
\affiliation{J. R. Macdonald Laboratory, Department of Physics, Kansas State University, Manhattan, Kansas 66506, USA}
\affiliation{Department of Physics, University of Connecticut, Storrs, Connecticut 06269, USA}

\author{Uwe Thumm}
\author{Artem Rudenko}
\affiliation{J. R. Macdonald Laboratory, Department of Physics, Kansas State University, Manhattan, Kansas 66506, USA}

\date{\today}

\begin{abstract}

The strong-field control of plasmonic nanosystems opens up new perspectives for nonlinear plasmonic spectroscopy and petahertz electronics. Questions, however, remain regarding the nature of nonlinear light-matter interactions at sub-wavelength spatial and ultrafast temporal scales. Addressing this challenge, we investigated the strong-field control of the plasmonic response of Au nanoshells with a SiO$_2$ core to an intense laser pulse. We show that the photoelectron energy spectrum from these core-shell nanoparticles displays a striking transition between the weak and strong-field regime. This observed transition agrees with the prediction of our modified Mie-theory simulation that incorporates the nonlinear dielectric nanoshell response. The demonstrated intensity-dependent optical control of the plasmonic response in prototypical core-shell nanoparticles paves the way towards ultrafast switching and opto-electronic signal modulation with more complex nanostructures.

\end{abstract}


\maketitle

The ability to reversibly manipulate the electronic structure and optical response of nanometer-sized materials has recently attracted substantial attention \cite{Stockman2011, Law2013,Krausz2014}. A hallmark property of nanostructures is the capacity to design and fabricate systems to take advantage of the tunable, size-, shape-, frequency-, and material-dependent properties as a means of tailoring specific optical responses. This holds the promise to both further our understanding of the transient electronic response in solid matter as well as enable new applications such as novel opto-electronics \cite{Krausz2014}, plasmonically enhanced light harvesting \cite{Sheldon2014}, and photocatalysis \cite{Photocatalysis,Wu2011}. Among different configurations, 
composite nanostructures, such as core-shell nanoparticles, consisting of a dielectric core and a thin metallic shell, are of special interest for their exceptionally large plasmonic field enhancements and high tunability of absorption spectra \cite{Li2018PRL,Halas2006}, generating novel applications in optical imaging and photothermal cancer therapy \cite{Rastinehad2019,Chen2014targeting}.  Precise control of the optical response, typically achieved by manipulating the geometric structures \cite{Halas2006}, is the key to utilizing their unique plasmonic properties.

Investigations into such optical properties in nanostructures have been conducted by studying their plasmonic response, in particular, their plasmonic near-field enhancement \cite{Li2018PRL,LiThesis,Hommelhoff2013Probing,Dombi2017Probing}. Photoelectrons provide an excellent window into understanding the dynamics of these interactions due to their sensitivity on the sub-wavelength spatial and ultrafast temporal scales. Photoelectron spectroscopy utilize these photoelectrons emitted during the interaction of a nanoparticle with an intense, femtosecond laser, allowing for the unraveling of the fundamental contributions to their acceleration, including enhanced near-fields, surface rescattering and charge interactions \cite{Sussmann2015,Powell19,Zherebtsov2011}.
Experiments revealed the fundamental light-matter interaction processes during the optical response and associated electron dynamics in selected nanosystems, consistent with theoretical modeling of the induced plasmonic field near the nanostructure surface in the linear-response approximation \cite{Li2016,Li2017,Zherebtsov2011, Rupp2019}. 

However, recent investigations of nanoscale thin films and metasufaces have revealed significant nonlinear (Kerr) effects \cite{Xiao2018,Nookala16}. These motivate the study of such nonlinear effects in the optical response in composite nanostructures consisting of similar thin layers (e.g. core-shell nanoparticles).

\begin{figure*}
    \includegraphics[width=1.00\linewidth]{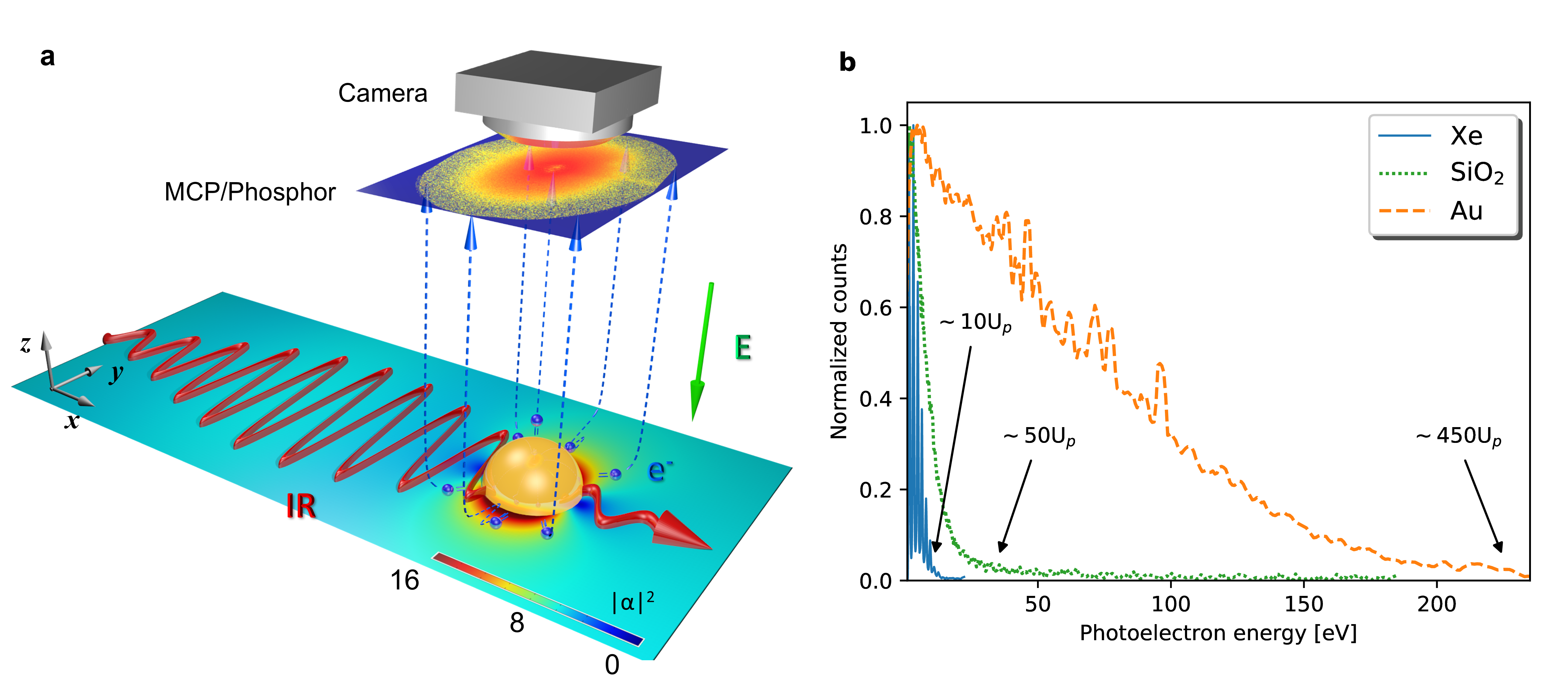}
    \caption{\textbf{Schematic of experiment and photoelectron spectra examples.} \textbf{a}, Schematic of an ultrafast laser interaction with a single, isolated nanoparticle in vacuum. The emitted photoelectrons are angle- and energy resolved using high-energy velocity map imaging (VMI) spectroscopy. \textbf{b}, Comparison of the obtained photoelectron energy distribution for photoemission from gaseous Xe (blue solid line), 120 nm diameter SiO$_2$ nanospheres (green dotted line), and 120 nm diameter Au nanospheres (gold dashed line), at a peak laser intensity of 8 TW/cm$^2$ at 780 nm wavelength. Respective cut-off energies in units of the ponderomotive energy $U_{\rm p}$ are indicated by arrows. All spectra are obtained from photoelectron counts within a 30$^\circ$ cone centered along the polarization direction.} \label{fig:1}
\end{figure*}  

Here, we investigate a type of core-shell nanoparticle specifically tailored to have a significant nonlinear effect which enables the intensity-dependent optical control of its plasmonic properties. This core-shell nanoparticle is comprised of a nanometer-thin gold coating encasing a larger dielectric silica core. We demonstrate, experimentally and theoretically, that as a function of the incident field intensity, a nonlinear response in the gold shell can be induced to control the plasmonic properties of the nanoparticle. At laser intensities below 0.1 TW/cm$^2$, the linear response dominates, resulting in a large plasmonic near-field. However, with increasing intensity, the onset of a nonlinear component of the complex index of refraction for gold decreases the skin depth and effectively reduces the magnitude of the near-field. This ability to manipulate the plasmonic properties of core-shell nanoparticles solely by tuning the external-field intensity substantiates a new method of precise control over the optical response in layered nanomaterials.

\section{Results and Discussion}

The photoelectron cut-off energy, defined as the highest observable electron energy, has been established as a gold standard for probing the induced plasmonic near-fields close to various nanostructure surfaces of different materials \cite{Hommelhoff2013Probing,Dombi2017Probing,Wang2020Cutoff,schotz2021onset}. This cut-off scales linearly with the cycle-averaged quiver energy of a free electron in a laser field, referred to as the "ponderomotive energy", $U_{\rm p} = e^2E^2/4m\omega^2 \propto I\lambda^2$ ($e$ and $m$ are electron charge and mass; $E$, $I$, $\omega$, and $\lambda$ designate the incident field strength, peak intensity, frequency, and wavelength of the laser pulse, respectively). Though near-fields are generally inhomogeneous (decreasing with distance), the fastest photoelectrons elastically rescatter and gain most of their kinetic energies near the nanoparticle surface \cite{Zherebtsov2011,SummersDissertation}, well within the typical spatial range of the near-field enhancement \cite{Li2018PRL,Sussmann2011,SummersDissertation}. Therefore, rescaling photoelectron cut-off energies with the incident-field $U_{\rm p}$ reveals information about the plasmonic near-field enhancements, independent of the incident-field intensity.

We determine the cut-off energies for single, isolated nanoparticles photoionized by femtosecond laser pulses, employing a high-energy velocity map imaging (VMI) spectrometer to measure the energy- and angle-resolved photoelectron spectra (see Methods for additional details). Figure \ref{fig:1}a illustrates a simplified schematic of the interaction and subsequent electron propagation and detection. Figure \ref{fig:1}b shows the comparison of typical photoelectron energy spectra and their respective cut-off energies for atomic Xe, SiO$_2$ nanospheres, and Au nanospheres, exemplifying the substantial increase in the kinetic energy of electrons emitted from nanoparticles.

To demonstrate the feasibility of observing the signature of the near-field in these nanosystems, we measured the size-dependent photoelectron cut-off energies for solid Au nanospheres. \frf{fig:2}a shows the cut-off energies for diameters ranging from 5 nm to 400 nm at several intensities. The plasmonic near-field response was investigated by rescaling the cut-off energies to their respective incident-field ponderomotive energy $U_{\rm p}$, as shown in \rf{fig:2}b. For comparison, the cut-off energies for SiO$_2$ nanospheres at similar diameters and intensities are also plotted \cite{Powell19}. The $U_{\rm p}$-rescaled cut-off energies for both materials are shown to be independent of the laser field intensity, within the size and intensity range of this work, as evidenced in \rf{fig:2}b by the overlap of the data points for each particular size. The cut-off energies for Au nanospheres indicates a prominent peak at diameter $D$ = 200 nm and more energetic photoelectrons than for SiO$_2$ nanoparticles, for which only a slight, monotonic increase with diameter occurs.

\begin{figure}
    \includegraphics[width=1.00\linewidth]{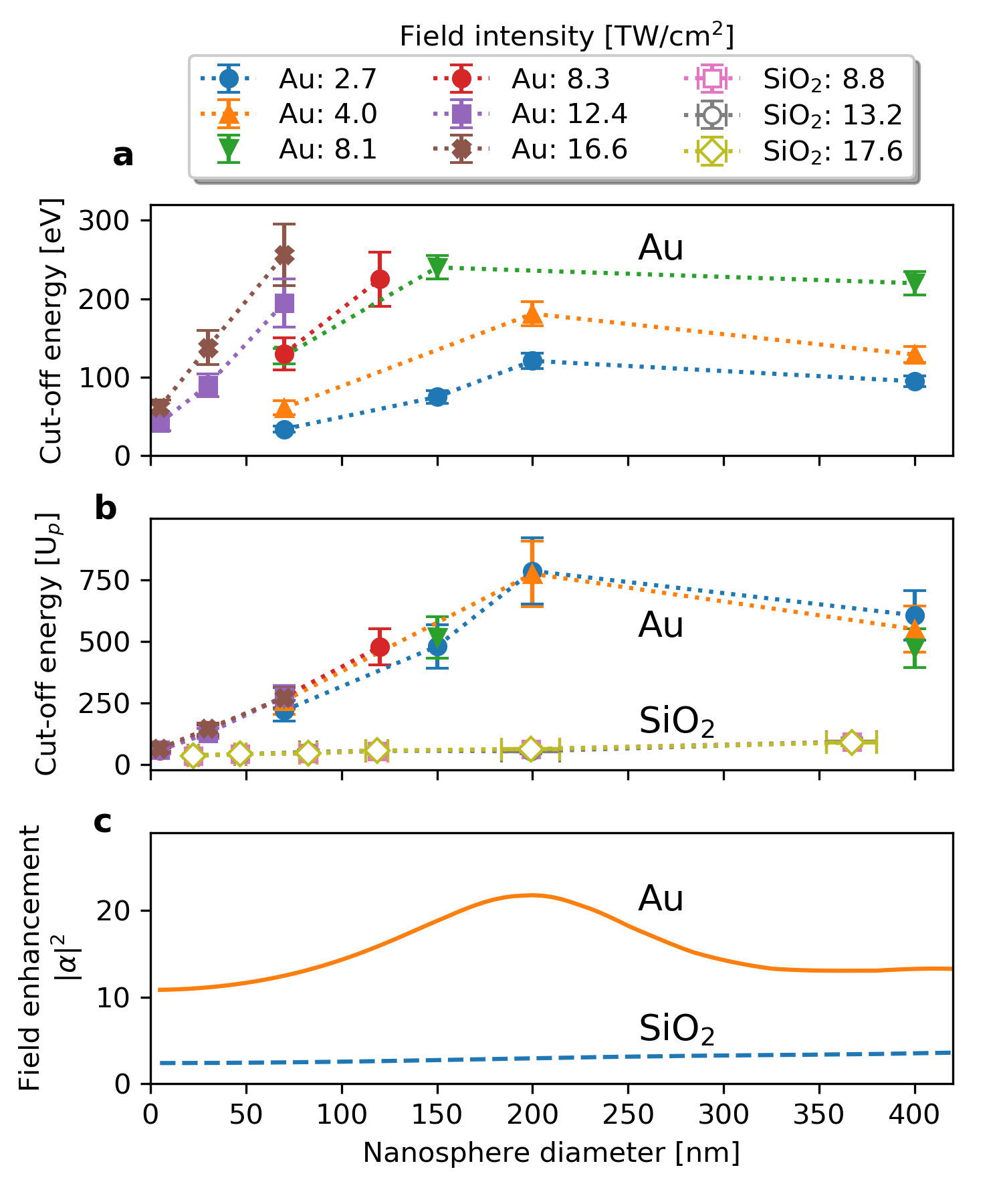}
    \caption{\textbf{Probing field enhancements with photoelectron cut-off energies.} \textbf{a}, Size-dependent maximum cut-off energies from experimental VMI spectra for various laser intensities for Au nanospheres in units of eV and for an incident pulse wavelength of 780 nm. \textbf{b}, The same cut-off energies as in \textbf{a} for Au and SiO$_2$ nanospheres rescaled to incident-field ponderomotive energy $U_{\rm p}$. \textbf{c}, Simulated near-field enhancement $|\alpha|^2$ for Au (orange solid line) and SiO$_2$ nanospheres (blue dashed line). } \label{fig:2}
   
\end{figure}

Further analysis in  \rf{fig:2}c reveals that the $U_{\rm p}$-rescaled cut-off energies resemble the maximum near-field intensity enhancement (which we refer to as ``field enhancement'' $|\alpha|^2$), in  \rf{fig:2}b. For the results shown in \rf{fig:2}c, we calculated the induced near-field by numerically solving the Mie equations \cite{Mie,Stratton} using indices of refraction within the linear optical response for Au and SiO$_2$ nanospheres \cite{Johnson&Christy}. The field enhancement is defined by the maximum total field intensity with respect to the incident intensity and typically located close to the particle surface \cite{Li2016,Li2017}. Though the exact dependence between the cut-off energy and field enhancement is determined by multiple effects (such as rescattering and Coulomb interactions) \cite{Sussmann2015,Zherebtsov2011}, \rf{fig:2}b and \rf{fig:2}c reveal that the cut-off energy does, however, accurately reflect the dominant signature of field enhancement profiles for Au and SiO$_2$ nanospheres (see Supplemental Note 2). This agreement supports our use of the cut-off energy as an indicator for observing changes in the magnitude of the plasmonic field enhancement.



\begin{figure}
    \includegraphics[width=1.00\linewidth]{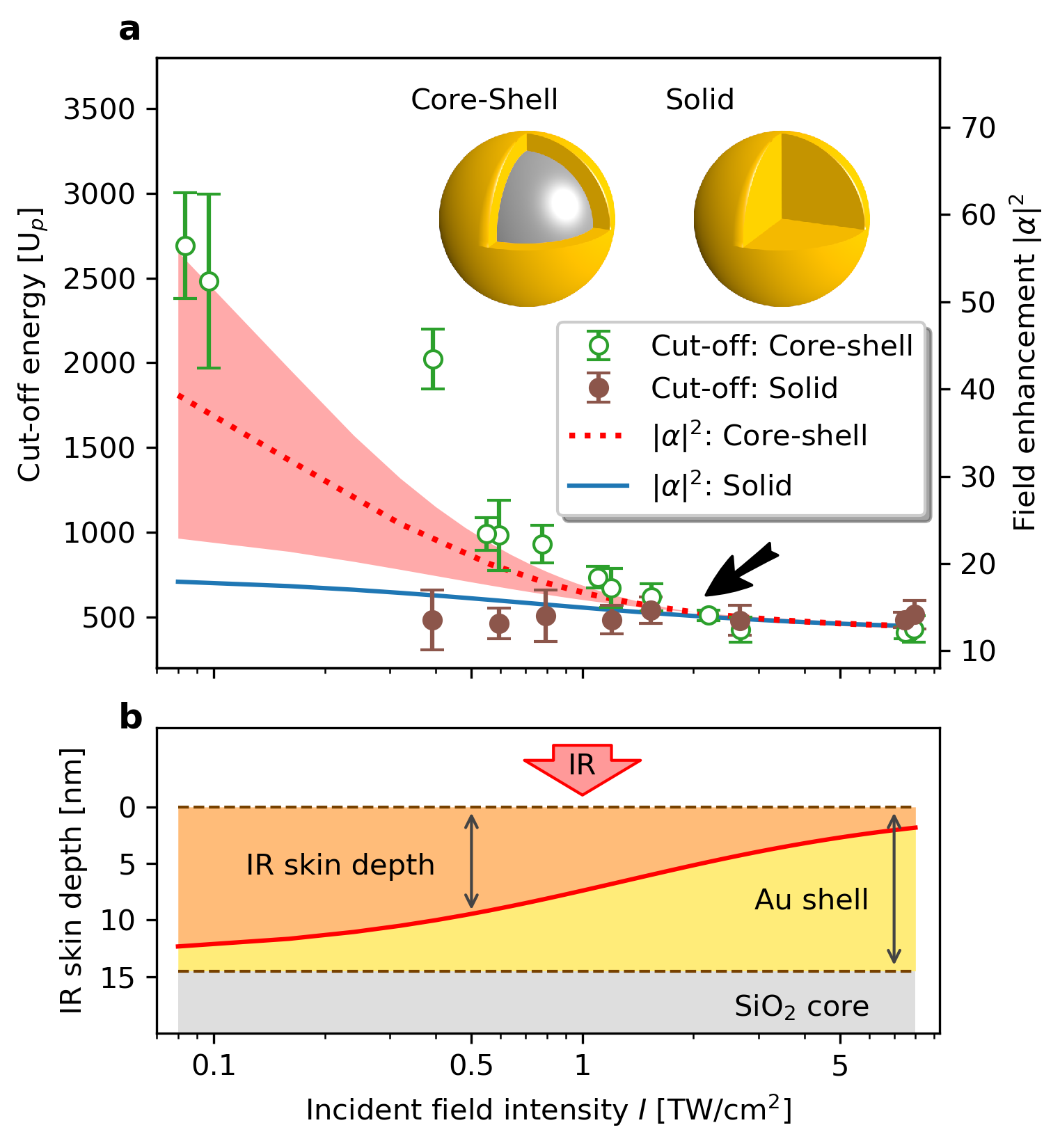}
    \caption{\textbf{Probing and analyzing intensity-dependent field enhancement of core-shell structures and solid Au nanospheres.} \textbf{a}, Maximum photoelectron cut-off energies in the units of incident-field ponderomotive energy $U_{\rm p}$, for SiO$_2$-core-Au-shell structures (hollow markers) and solid Au nanospheres (solid markers), as a function of the incident-field intensity $I$. The core-shell structures have an outer diameter $D_2 = 147 \pm 7$ nm and an inner diameter $D_1 = 118 \pm 4$ nm, while the solid Au nanospheres have a diameter of $150 \pm 5$ nm. The simulated field enhancement $|\alpha|^2$ (scale on the right) for core-shell structures (red dotted line) and solid Au nanospheres (blue solid line) are also plotted for comparison. The red shaded area is the calculated uncertainty range caused by the manufacturing dispersity of the inner and outer radii of the core-shell structures (See Supplementary Note 4). The black arrow indicates the selected convergence of the core-shell and solid nanospheres in cut-off energies and in field enhancements. \textbf{b}, Calculated IR skin depth (red line and darkened area) for a Au shell thickness of 14.5 nm, as a function of the incident-field intensity $I$. All results are for an incident pulse wavelength of 780 nm.} \label{fig:3}
\end{figure}

To investigate the strong-field control of the plasmonic properties in a layered nanostructure, we examined SiO$_2$-core-Au-shell structures. Core-shell nanoparticles were specifically chosen to elucidate any intensity-related changes to the field enhancement due to their unique configuration in comparison with solid Au nanospheres of a comparable size. The nanometer-thin Au shell surrounding a dielectric core results in a stronger localization of the plasmonic near-field when compared to solid Au nanospheres. We neglect any laser-induced propagation effects as the nanoparticle radius is much smaller than the incident pulse wavelength.
\frf{fig:3}a shows our measured intensity-dependent photoelectron cut-off energies, rescaled with the incident-field ponderomotive energy $U_{\rm p}$ at 780 nm wavelength, for core-shell structures and solid Au nanospheres of approximately 150 nm diameter.

The $U_{\rm p}$-rescaled cut-off energy for solid Au nanospheres remains approximately unchanged ($\sim$ 500 $U_{\rm p}$) for different incident-field intensities, indicating that the field enhancement near solid Au nanospheres is nearly independent of the intensity  (\rf{fig:3}a). In contrast, the cut-off energy for the core-shell structures varies drastically within the intensity range sampled. At low intensities, this energy is large when scaled by $U_{\rm p}$ (2000 $\sim$ 3000 $U_{\rm p}$, approximately 10 eV). However, it rapidly decreases and converges to the nearly identical value of the solid Au nanospheres beyond $\sim$ 2 TW/cm$^2$, as indicated by the black arrow in \rf{fig:3}a. This implies that the field enhancement of the core-shell structure does not remain constant with the increasing laser intensity, but rather begins (at low intensity) at a value significantly higher for solid Au nanospheres, before quickly decreasing to a similar value at higher intensities. The contrast of these two particles is particularly strong at very low intensities ($\sim$ 0.1 TW/cm$^2$), where no photoemission is observed for solid Au nanospheres due to the extremely weak electromagnetic field. A measurable amount of photoelectrons with over 2000 $U_{\rm p}$ cut-off energy are still observed from core-shell structures, further confirming that a large field enhancement is induced at low intensities.


The nonlinear optical response of the Au shell is the key to understanding the observed effects in core-shell nanoparticles. Note that only the linear optical response is included in \rf{fig:2}c. To introduce the nonlinear response, we apply a simple and widely used model to account for intensity-dependent changes in the index of refraction for Au \cite{BoydBook},
\begin{equation}\label{eq:n}
n = n_0 + n_2I,
\end{equation}
where $n_0$, an experimentally determined complex number \cite{Johnson&Christy}, is the linear index of refraction employed for our simulation results in \rf {fig:2}. $n_2$ is related to the third-order susceptibility $\chi^{(3)}$ (or Kerr effect) \cite{Boyd2004},
\begin{equation}\label{eq:n2}
n_2(\text{m}^2/\text{W}) = \frac{283}{n_0 \Re(n_0)}\chi^{(3)}(\text{m}^2/\text{V}^2).
\end{equation}
$\Re(n_0)$ is the real part of $n_0$ and $\chi^{(3)} = (-9.1+0.35i)\times10^{-19}~\text{m}^2/\text{V}^2$, according to a measurement using 15 nm Au films \cite{Liu2016}, close to the Au shell thickness in our work. At 780 nm wavelength, $n_2 = (0.026+3.65i) \times 10^{-12} (\text{W}/\text{cm}^2)^{-1}$ is predominantly imaginary.
When $I \rightarrow 0$, the linear optical response dominates and $n \approx n_0$. As the intensity increases, the imaginary part $\Im(n) = \Im(n_0) + \Im(n_2)*I$ increases, and the nonlinear effect starts to emerge.

We estimate the normal-incident IR skin depth $\sigma$ following \cite{Hecht2002} as,
\begin{equation}\label{eq:skinDepth}
	\sigma = \frac{c}{2\omega\Im(n)} = \frac{c}{2\omega \big[ \Im(n_0) + \Im(n_2)*I \big]},
\end{equation}
where $c$ is the speed of light. 
Thus, the IR skin depth decreases with incident-field intensity in this particular situation. \frf{fig:3}b plots the intensity-dependent IR skin depth, in comparison with the Au shell thickness ($\sim$ 14.5 nm) of the core-shell structures. At low intensities (< 0.1 TW/cm$^2$), the skin depth is approximately 13 nm, comparable to the Au shell thickness, suggesting a considerable amount of the IR field reaches the SiO$_2$ core. Since the optical response of a core-shell nanoparticle is extremely sensitive to the fields at both the inner and outer surfaces, their larger field enhancement results from the penetration of the external field into the Au-SiO$_2$ interface. This effect is incorporated by applying the boundary conditions at both interfaces when solving the Mie equations \cite{Li2018PRL}. However, as the intensity increases, $\sigma$ rapidly drops and approaches 2 nm at 8 TW/cm$^2$, well below the Au-shell thickness, preventing the IR field from penetrating the Au shell, i.e., shielding the SiO$_2$ core. Therefore, the outer Au surface becomes the dominant factor for determining the optical response (and, thus, the field enhancement), causing the core-shell structures to appear to be indistinguishable from solid Au nanospheres of the same outer diameter.

Note that, while a recent study included the nonlinear response of SiO$_2$ nanoparticles \cite{Rupp2019}, we neglect such effects in the SiO$_2$ core of the core-shell structures, since the field near the Au-SiO$_2$ interface is too heavily dampened (not exceeding 10$^{10}$ W/cm$^2$) to induce any significant nonlinear response in this study.  

\frf{fig:3}a also plots the Mie-simulated field enhancement $|\alpha|^2$ for core-shell structures and solid Au nanospheres, including the nonlinear optical response. The field enhancement of solid Au nanospheres only decreases slightly from 18 to 13 as the intensity increases. This indicates that the nonlinear response, while present in solid Au nanospheres, is insignificant and does not induce a measurable difference exceeding experimental uncertainty in this work, justifying our calculation of the results in \rf{fig:2} using a linear-response approximation. The core-shell structures, on the other hand, start with a field enhancement up to 60 at low intensities, where linear response dominates. This large enhancement is responsible for the photoemission observed from core-shell structures at low intensities, which is absent in solid Au nanospheres, as well as the significantly larger cut-off energy. As the intensity increases, the nonlinear response results in a decreasing skin depth (\rf {fig:3}b), leading to significantly smaller field enhancements that eventually converge with our results for the solid Au nanospheres beyond $\sim$ 2 TW/cm$^2$. The profiles of the simulated field enhancements, especially the convergence at $\sim$ 2 TW/cm$^2$, are in excellent agreement with that of the measured cut-off energies for both nanoparticles. This success validates the use of the simple model in \re{eq:n} and shows that the laser intensity has a significant impact on the plasmonic response of core-shell structures, in contrast to their solid Au counterparts.

\begin{figure}
	\includegraphics[width=1.00\linewidth]{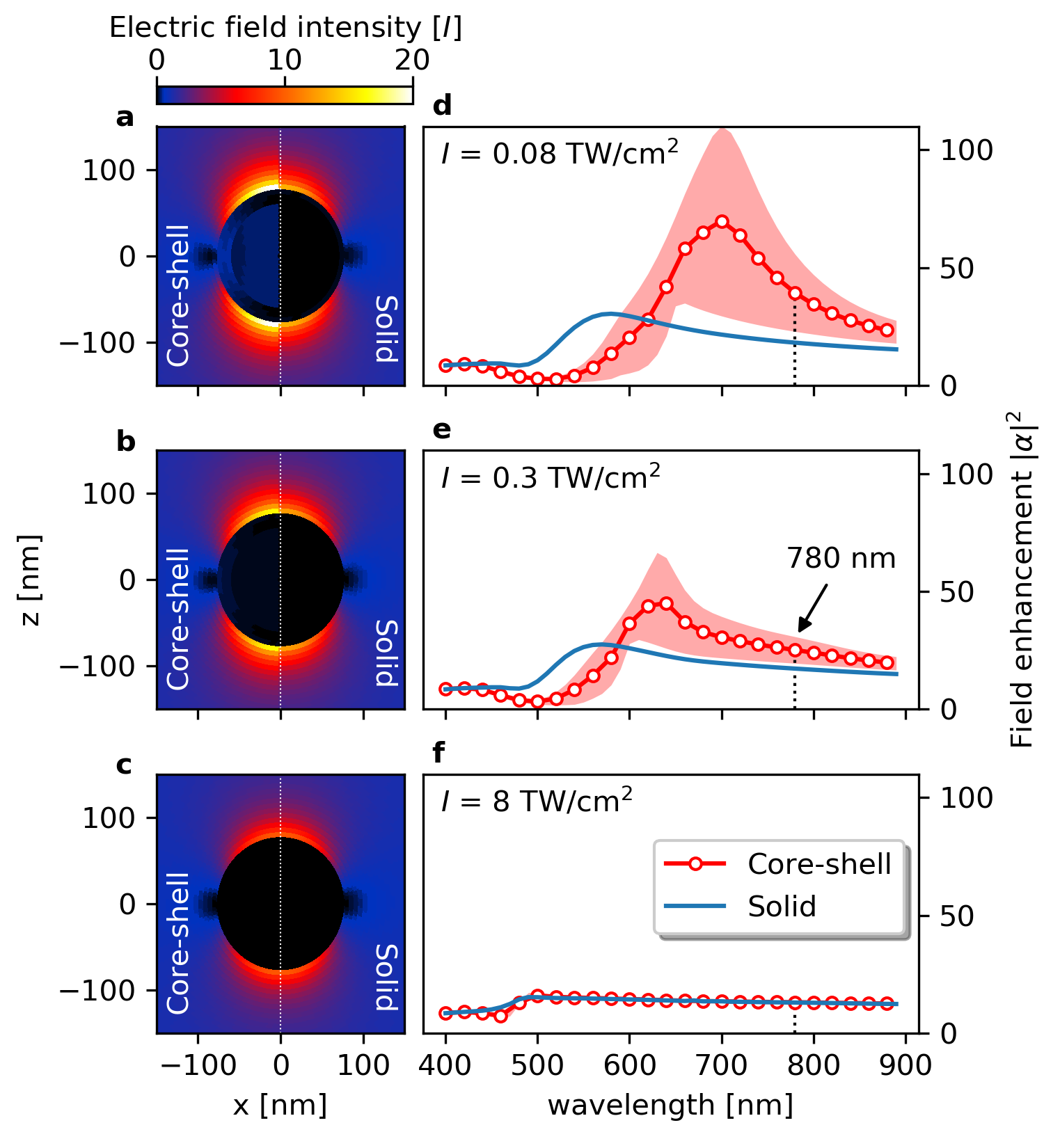}
	\caption{\textbf{Controlling plasmonic properties of core-shell structures.} \textbf{a-c}, Simulated inhomogeneous electric-field intensity distribution for core-shell structures (left half) and solid Au nanospheres (right half), for the incident-field wavelength of 780 nm and intensities of (\textbf{a}) 0.08, (\textbf{b}) 0.3, and (\textbf{c}) 8 TW/cm$^2$. \textbf{d-f}, Simulated field enhancement $|\alpha|^2$ as a function of incident wavelength, under the same respective incident-field intensities, for core-shell (red circled lines) and solid Au nanospheres (blue solid lines). The red shaded areas are the uncertainty range caused by the manufacturing dispersity of the inner and outer radii of the core-shell structures.} \label{fig:4}
\end{figure}


A more in-depth understanding of such an impact and the broader implications can be achieved by extending our line of investigation to a larger spectral range. Figures \ref{fig:4}d-f show the calculated field enhancement of core-shell structures and solid Au nanospheres as a function of the incident-field wavelength for incident-field intensities of 0.08, 0.3, and 8 TW/cm$^2$, respectively. For each corresponding incident-field intensity, Figs. \ref{fig:4}a-c compare a section of our simulated electric-field-intensity distribution between the two types of nanoparticles for an incident field at 780 nm wavelength. 

At 0.08 TW/cm$^2$, where the linear optical response dominates, a weak yet noticeable field penetrates the Au shell of the core-shell structure. The propagation of the external field across the Au-SiO$_2$ interface yields the significantly more intense and red-shifted resonant spectra, as compared with solid Au nanoparticles. Such pronounced differences show that the plasmonic response of core-shell structures, including the absorption resonance and the near-field magnitude, can be tuned by changing parameters in the production of the nanospheres (core diameter, shell thickness and composition) \cite{Li2018PRL,Halas2006,Rastinehad2019,Chen2014targeting}.

As the incident-field intensity increases, due to the growing nonlinear response of the Au shell preventing penetration of the field, the resonance feature for the core-shell structures starts to blue shift and decrease in magnitude, before gradually becoming indistinguishable to solid Au nanospheres across the spectra. We thus demonstrate that for core-shell nanoparticles, the tunable plasmonic response can be effectively switched ``on'' and ``off'' by simply controlling the external field intensity. For intensities well below a threshold ($\sim$ 2 TW/cm$^2$ in this work), the tunable plasmonic response of core-shell structures is switched ``on'', manifesting a large and red-shifted resonance. For intensities above the threshold, the response is effectively switched ``off'', and the core-shell structures appear to be indistinguishable from the solid Au nanospheres. According to \re{eq:skinDepth}, for a chosen material, such a threshold is determined primarily by the shell thickness, where a thinner shell requires a larger intensity threshold to ensure the IR skin depth is smaller than the shell thickness. This shows core-shell nanoparticles can be carefully synthesized to have a designated intensity threshold so that their plasmonic properties are controllable by manipulating the external field intensity.

\section{Conclusions}

We have demonstrated the ability to control the plasmonic response of a layered nanostructure solely by varying the laser intensity. This was accomplished by measuring the photoelectron cut-off energies from single, isolated core-shell nanoparticles and using them as a sensitive probe of the plasmonic field. Experimental signatures of a non-constant, intensity-dependent near-field were verified by a modified Mie theory as the direct result of the nonlinear optical response of the outer gold shell. Further analysis revealed that the decreasing skin depth into the nanoparticle surface at laser intensities above $\sim$ 2 TW/cm$^2$ effectively shields the SiO$_2$ core, rendering the magnitude of its near-field identical to that of a solid Au nanosphere. These results suggest a new intensity-dependent strong-field control of the plasmonic response in layered nanostructures. While such responses in layered nanostructures are known to be tunable by their physical structure, we demonstrated that they can further be effectively switched ``on'' and ``off'' solely by controlling the external-field intensity.

This intensity-dependent optical control of the plasmonic response could hold the keys to new lines of research and implementations based on layered nanostructures. For example, many applications require nanosystems such as core-shell structures to be tuned to precise resonant wavelengths \cite{Rastinehad2019,Chen2014targeting}. Our work unlocks a new tuning mechanism where a single core-shell structure can be manipulated to have an adjustable resonance and optical properties, dependent merely on the applied laser intensity.
In other applications, such as photocatalysis \cite{Wu2011} and field-induced molecular reactions \cite{Rosenberger2020}, core-shell structures could be versatile substitutes for the currently used solid nanoparticles. Not only can they be tailored to provide larger field enhancements and substantially reduce the laser-intensity requirements, but such large enhancements can also be automatically turned ``off'' by the nonlinear response to avoid overexposure at high intensities. This can have significant impacts in areas such as metamaterials, plasmonics and opto-electronics.

\bibliographystyle{naturemag}
\bibliography{NL}

\section{Methods}

\subsection{Experimental setup}

The laser setup and electron detection apparatus at the James R. Macdonald Laboratory at Kansas State University are described in more detail in \cite{Powell19}. Briefly, the experiments used a Ti:Sapphire-based chirped pulse amplification (CPA) system generating 25 fs pulses at 780 nm central wavelength. Photoelectron spectra were captured in a thick-lens, high-energy velocity map imaging (VMI) spectrometer \cite{Kling2014} capable of gathering up to 350 eV electron energy. The custom nanoparticle source produces a continuous beam of nanoparticles into vacuum. Spherical nanoparticle samples were selected for their narrow size distribution and overall purity. The initial nanoparticle concentration was also carefully chosen to avoid the formation of clusters in the nanoparticle beam.

\subsection{Laser intensity characterization}

The peak laser intensity was determined by analyzing the above-threshold (ATI) photoelectron energy distribution of atomic Xe with the aforementioned VMI under similar experimental parameters. The ponderomotive shift of the Xe ATI comb was measured as a function of the input-laser pulse energy, in order to derive the ponderomotive energy, $U_{\rm p} \propto I\lambda^2$, and thus, the peak laser intensity $I$ \cite{SummersDissertation, PowellThesis}. For intensities below the ionization threshold of Xe, the ratio of the pulse energies was used to extrapolate the peak intensity. See Supplementary Note 1 for more detail.

\subsection{Photoelectron cut-off determination}

The nanoparticle photolectron cut-off energy was extracted from the experimental VMI images in a method described in previous work \cite{Powell19, PowellThesis}. The detected elastically back-scattered photoelectrons are obtained from the non-inverted VMI images, for which the upper energy boundaries of the full 3D momentum sphere and the 2D projection are essentially the same. A radial distribution of these projections along the polarization direction accurately determines the maximum photoelectron energy.

\subsection{Mie simulations}

We simulated laser-induced plasmonic near-fields by solving the Mie equations \cite{Mie,Stratton} for plane waves scattered by spherical objects. The linear terms of the dielectric response of Au and SiO$_2$ enters as the frequency-dependent complex-valued index of refraction obtained from experiments \cite{Johnson&Christy,Palik}. Our inclusion of nonlinear effects is discussed in the text. The original Mie equations apply only to the solid spheres. For the simulations of core-shell structures, we extended the traditional Mie theory as outlined in Supplementary Note 3.

\section{Acknowledgment}

This work was supported by the Air Force Office of Scientific Research under award number FA9550-17-1-0369. J.L., E.S. and U.T. acknowledge the support by the NSF Grant No. 1802085. A.S. and D.R. were supported by the Chemical Sciences, Geosciences, and Biosciences Division, Office of Basic Energy Sciences, Office of Science, U. S. Department of Energy under Award No. DEFG02-86ER13491, which also covered laser operational costs. M.F.K acknowledges support by the German Research Foundation (DFG) via SPP1840, and by the European Research Council (ERC) via the FETopen project PetaCOM.



\section{Author contributions}

J.P. and J.L. contributed equally to this work. J.P., J.L., A.S., U.T. and A.R. conceptualized and conducted the study. J.P., A.S., S.J.R, M.D. and P.R. contributed to performing the experiment. J.P. and A.S. performed the data analysis. J.L. and U.T. developed the theoretical model, performed the simulations and interpreted the results. J.P., J.L., A.S., E.S., C.M.S., D.R., M.F.K., C.T.H., U.T. and A.R. discussed the results. J.L., J.P. and A.S. wrote the initial manuscript, which was revised with input from all authors.

\section{Additional information}    

The authors declare no competing financial interests.

\end{document}


\title{Supplementary Information}

\author{Jeffrey Powell}
\thanks{These authors contributed equally to this work.}
\affiliation{J. R. Macdonald Laboratory, Department of Physics, Kansas State University, Manhattan, Kansas 66506, USA}
\affiliation{Department of Physics, University of Connecticut, Storrs, Connecticut 06269, USA}
\affiliation{INRS, {\'E}nergie, Mat{\'e}riaux et T{\'e}l{\'e}communications, 1650 Bld. Lionel Boulet, Varennes, Qu{\'e}bec, J3X1S2, Canada}

\author{Jianxiong Li}
\thanks{These authors contributed equally to this work.}
\affiliation{J. R. Macdonald Laboratory, Department of Physics, Kansas State University, Manhattan, Kansas 66506, USA}
\affiliation{Department of Physics and Astronomy, Louisiana State University, Baton Rouge, Louisiana 70803, USA }

\author{Adam Summers}
\affiliation{J. R. Macdonald Laboratory, Department of Physics, Kansas State University, Manhattan, Kansas 66506, USA}
\affiliation{ICFO - Institut de Ciencies Fotoniques, The Barcelona Institute of Science and Technology, 08860 Castelldefels (Barcelona), Spain}

\author{Seyyed Javad Robatjazi}
\affiliation{J. R. Macdonald Laboratory, Department of Physics, Kansas State University, Manhattan, Kansas 66506, USA}

\author{Michael Davino}
\affiliation{Department of Physics, University of Connecticut, Storrs, Connecticut 06269, USA}

\author{Philipp Rupp}
\affiliation{Physics Department, Ludwig-Maximilians-Universit\"at Munich, D-85748 Garching, Germany}

\author{Erfan Saydanzad}
\affiliation{J. R. Macdonald Laboratory, Department of Physics, Kansas State University, Manhattan, Kansas 66506, USA}

\author{Christopher M. Sorensen}
\author{Daniel Rolles}
\affiliation{J. R. Macdonald Laboratory, Department of Physics, Kansas State University, Manhattan, Kansas 66506, USA}

\author{Matthias F. Kling}
\affiliation{Physics Department, Ludwig-Maximilians-Universit\"at Munich, D-85748 Garching, Germany}
\affiliation{Max Planck Institute of Quantum Optics, D-85748 Garching, Germany}

\author{Carlos Trallero-Herrero}
\affiliation{J. R. Macdonald Laboratory, Department of Physics, Kansas State University, Manhattan, Kansas 66506, USA}
\affiliation{Department of Physics, University of Connecticut, Storrs, Connecticut 06269, USA}

\author{Uwe Thumm}
\author{Artem Rudenko}
\affiliation{J. R. Macdonald Laboratory, Department of Physics, Kansas State University, Manhattan, Kansas 66506, USA}

\date{\today}

\pacs{}

\maketitle
\clearpage

\begin{figure}
    \includegraphics[width=1.00\linewidth]{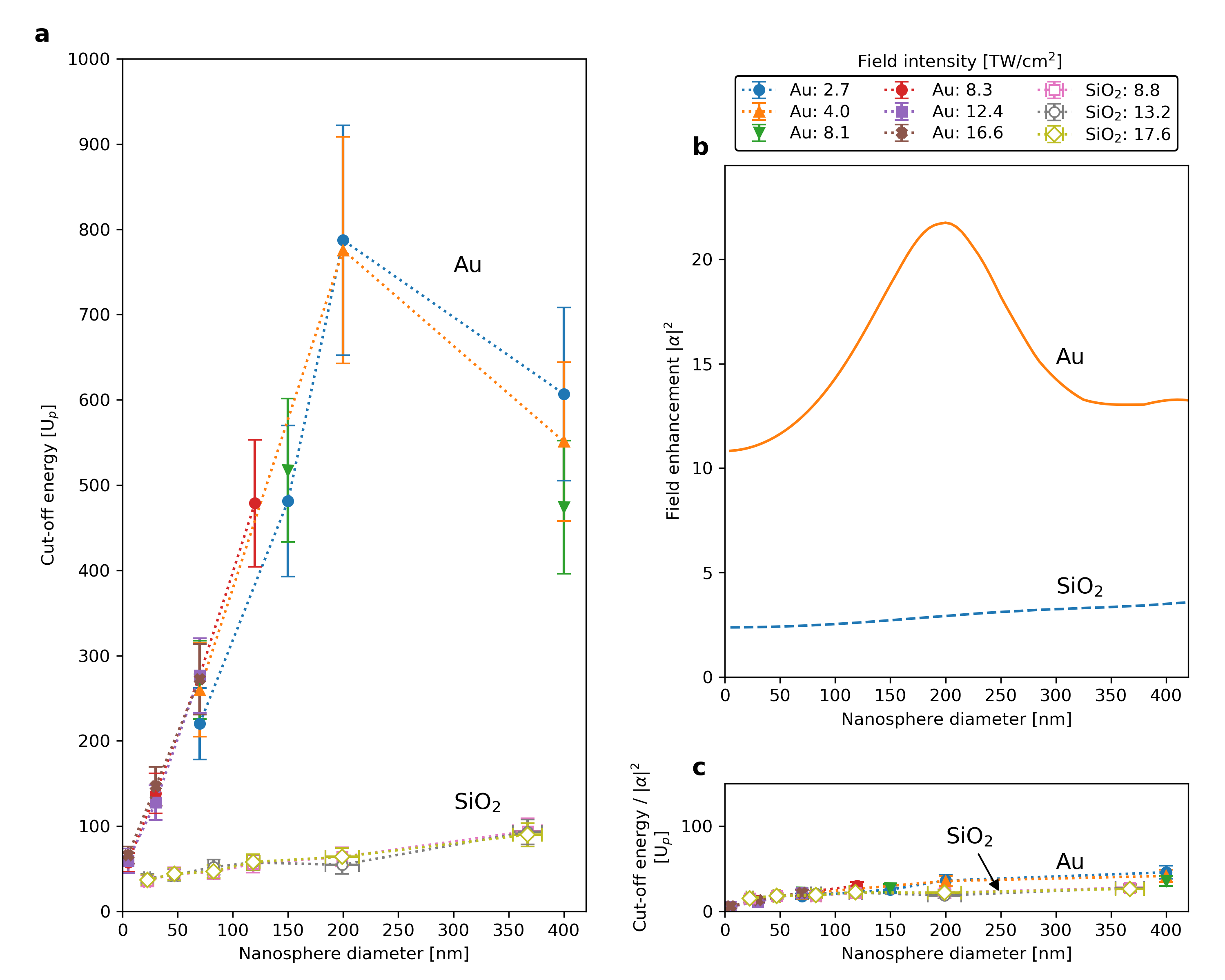}
    \caption{
        \textbf{a}, Size-dependent cut-off energies (re-scaled to the incident-field ponderomotive energy $U_{\rm p}$) from experimental VMI spectra for Au and SiO$_2$ nanospheres, various laser intensities, and a laser wavelength $\lambda=780$ nm. \textbf{b}, The Mie-simulated field enhancement $|\alpha|^2$ for Au (orange solid line) and SiO$_2$ nanospheres (blue dashed line). \textbf{c}, The same cut-off energies as in \textbf{a} for Au and SiO$_2$ nanospheres, shown on the same scale as in \textbf{a}, but divided by $|\alpha|^2$ to exclude the contribution from field enhancement.} \label{fig:SI1}
\end{figure}
\clearpage

\begin{figure}
    \includegraphics[width=1.00\linewidth]{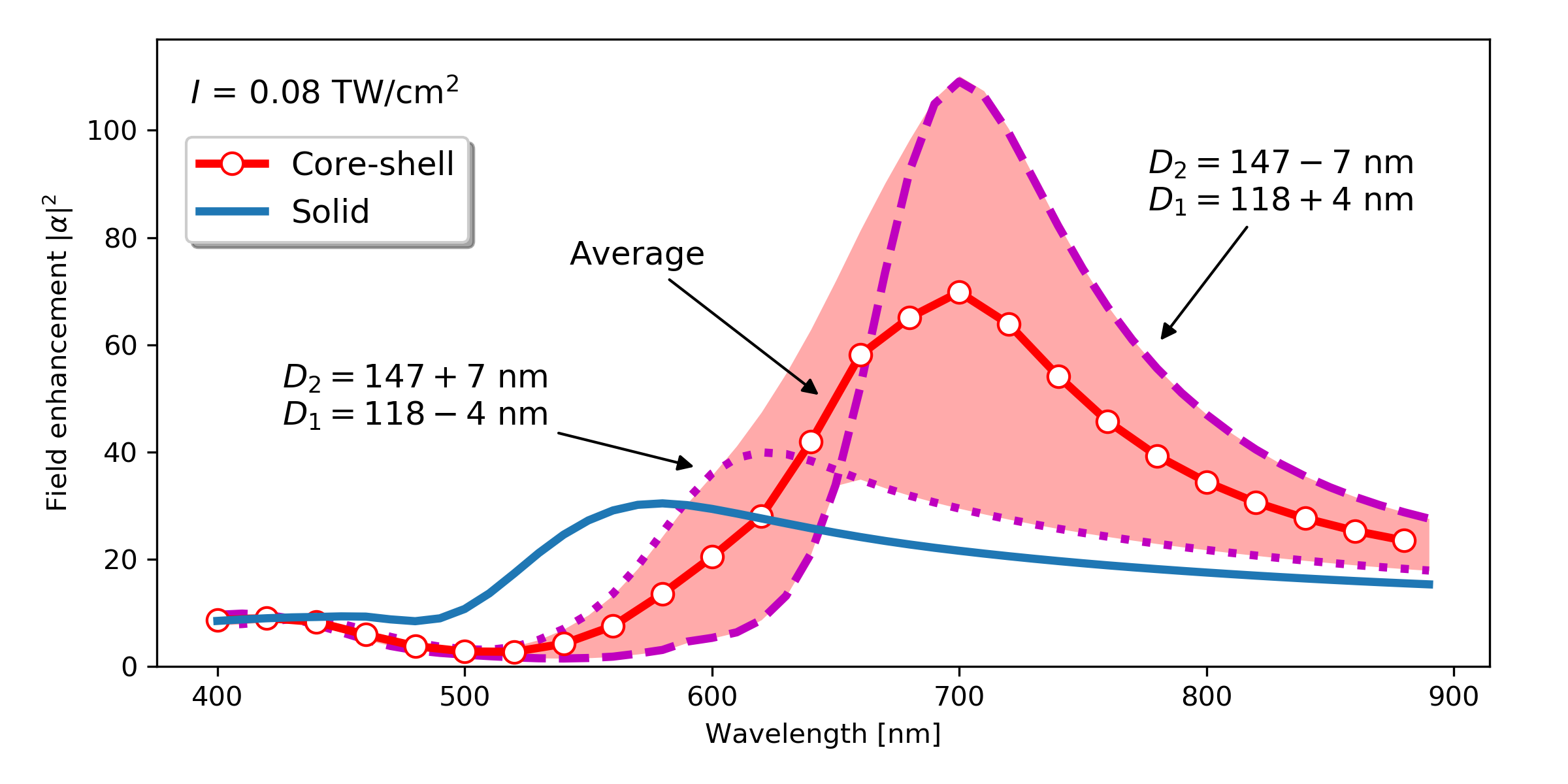}
    \caption{
        Illustration of the field-enhancement uncertainty as a result of manufacturing size dispersity (re-plot of \rf{fig:4}d). The core-shell structure has an outer diameter of $D_2 = 147 \pm 7$ nm and an inner diameter of $D_1 = 118 \pm 4$ nm. The purple dashed line is the field enhancement $|\alpha|^2$ simulated for $D_2 = 147 - 7$ nm and $D_1 = 118 + 4$ nm (thinnest shell). The purple dotted line is simulated for $D_2 = 147 + 7$ nm and $D_1 = 118 - 4$ nm (thickest shell). The red shaded area is the uncertainty range simulated with all possible combinations of outer and inner radii within the manufacturing dispersity. The red circled line is the average value within such uncertainty at each wavelength. The field enhancement of the solid Au nanospheres of the same size is shown for comparison as the blue solid line. The intensity of the simulation is $I$ = 0.08 TW/cm$^2$.} \label{fig:SI2}
\end{figure}
\clearpage

\section{Supplementary Note 1: Experimental setup}

The experimental setup for imaging and measuring photoelectron emission made use of a thick-lens, high-energy velocity map imaging (VMI) spectrometer as described in \cite{Powell19}. Laser pulses with a central wavelength of $\lambda=780$ nm and pulse duration of 25 fs were focused into the nanoparticle beam at the center of the interaction region of the VMI spectrometer. The intensity of the laser field was adjusted by a $\lambda/2$ waveplate in combination with a linear polarizer while the polarization was in the plane of the detector. The emitted electrons were projected onto a paired microchannel plate (MCP) and phosphor screen (P47 type), and the resulting images were recorded by a CMOS camera (Mikrotron EoSens 3CXP). The MCP was gated (200 - 400 ns) on the arrival time of the electrons using a high-voltage switch which effectively reduced the background contribution. The use of a real-time hit-finder software routine allowed for single-shot photoelectron spectra to be collected at greater than 1 kHz. The nanoparticle spectra were then analyzed by taking a radial projection along the polarization direction to determine the photoelectron cut-off.

The nanoparticle source delivers a continuous beam of single, isolated nanoparticles to the interaction volume in the direction perpendicular to the laser propagation. The source aerosolizes a colloid of nanoparticles that are subsequently dried by a solid-state counter-flow membrane dryer to remove the solvent (water) from the carrier gas (N$_2$). Additional care was taken to minimize the probability of clusters in the nanoparticle beam to prevent an amalgamation of photoelectrons from monomers and clusters \cite{Rosenberger2020} by keeping the number density below $10^{10}$ nanoparticles per mL. The in-vacuum beam density was increased using an aerodynamic lens system to focus the gas-phase nanoparticles while a three-stage differentially pumped arrangement removed excess carrier gas in vacuum. The typical pressure in the experimental chamber was $10^{-6}$ Torr with the nanoparticle source running, while the background pressure was  $10^{-8}$ Torr.

Spherical nanoparticle samples were selected for their narrow size distribution (<12\%), solvent choice and overall purity from commercially available sources. Solid gold nanoparticles (Cytodiagnostics, Inc) were stabilized with citrate while gold nanoshells and SiO$_2$ nanospheres (nanoComposix, Inc) were coated with polyethylene glycol (PEG, 5kDa) and silanol, respectively.

To measure the laser intensity and calibrate the VMI, above-threshold ionization (ATI) in Xe was measured at the same laser parameters as for the nanoparticles. The ponderomotive shift  of the Xe ATI comb was measured as a function of the input laser pulse energy in order to derive the ponderomotive energy,  $U_{\rm p} = e^2E^2/4m\omega^2 \propto I\lambda^2$, and thus, the peak laser intensity $I$. For experimental intensities below the ionization threshold of Xe, the ratio of the pulse energies was used to extrapolate the peak intensity. The spacing of consecutive Xe ATI peaks corresponds to the central incident photon energy and was used for energy calibration of the VMI \cite{Powell19}.

\section{Supplementary Note 2: Further discussion about the relation between cut-off energy and field enhancement}

\frf{fig:2} in the main text illustrates the relation between  $U_{\rm p}$-scaled cut-off energies and simulated field enhancements. It supports our use of the cut-off energy as an indicator for observing changes in field enhancement. A more detailed illustration can be seen in \rf{fig:SI1}. 

Figures \ref{fig:SI1}a-b are plotted with the same data set as in Figs. \ref{fig:2}b-c in the main text, respectively. \frf{fig:SI1}c shows the same cut-off energies as \rf{fig:2}a, but divided by $|\alpha|^2$, to scale out the contribution from field enhancement. Both \rf{fig:SI1}a and \rf{fig:SI1}c are plotted on the same vertical scale as a comparison. 

By scaling out the field enhancement contribution, \rf{fig:SI1}c shows much lower and nearly flat curves for both Au and SiO$_2$ nanospheres, as compared to \rf{fig:SI1}a. In addition, the differences between two materials are significantly reduced. Even though previous research showed that the exact dependence between the cut-off energy and field enhancement is convoluted due to multiple contributing factors \cite{Sussmann2015,Zherebtsov2011}, \rf{fig:SI1}c shows that field enhancement is the main contributor governing the changes in cut-off energies, within the range of our experimental parameters. This allows us to use the cut-off energy as an indicator for changes in the field enhancement.

It is also worth mentioning that this relation is seemingly less accurate for small nanoparticles by comparing \rf{fig:SI1}a and \rf{fig:SI1}b. However, \rf{fig:SI1}c still shows good convergence at small sizes due to the fact that the photoelectron yield decreases as the nanoparticles size decreases \cite{Zherebtsov2011}. Since electron-electron interaction is reportedly a significant component in the photoelectron dynamics \cite{Zherebtsov2011}, a decreasing photoelectron yield causes a decrease in the overall electron-electron interaction and, ultimately, the cut-off energies for both small Au and SiO$_2$ nanospheres, reducing the difference in $U_{\rm p}$ by directly comparing \rf{fig:2}a and \rf{fig:2}b.

We compared the core-shell structures and solid Au nanospheres of the same outer diameter ($\sim$ 150 nm). Therefore, the observed differences in cut-off energies are mainly due to the differences in the field enhancements.

\section{Supplementary Note 3: Modified Mie equations for core-shell structures}

Mie theory is a well established theory to simulate the electromagnetic fields of plane waves scattered by spherical objects \cite{Mie,Stratton}. The original Mie theory is restricted to solid spherical objects with a single spatially homogeneous index of refraction. This theory is sufficient to simulate the near fields of solid Au and SiO$_2$ nanospheres, as produced in \rf{fig:2} in the main text. Simulating the scattering fields for core-shell structures, however, requires its extension. Taking advantage of the spherical symmetry, these modifications can be included analytically as follows.

Suppose an incident plane wave with frequency $\omega$ travels in vacuum (index of refraction $n_0=1$) along the positive $\zu$-axis, with the electric component polarized along the $\xu$-axis. Its electric field is given by,
\begin{equation}
    \bE_i\brt = \xu E_0 e^{ik_0z-i\omega t},
\end{equation}
where $E_0$ is the incident-field amplitude, the subscript ``$0$'' stands for vacuum, and ``$i$'' stands for ``$incident$''. We consider a spherical core-shell nanoparticle centered at the origin of our coordinate system, with inner and outer radii $R_1=D_1/2$ and $R_2=D_2/2$, respectively. The core and shell materials are assumed to have homogeneous, complex-valued indices of refraction of $n_1$ and $n_2$, respectively. For convenience, we neglect magnetic materials in this discussion, and set permeability of all materials to 1. 

The total electro-magnetic field can now be expressed as,
\begin{equation}\label{eq:total}
    \bE\brt = 
    \begin{cases}
        \bE_i\brt+\bE_r\brt  \\
        \bE_s\brt  \\
        \bE_t\brt
    \end{cases} 
    \quad,\quad\quad
    \bH\brt = 
    \begin{cases}
        \bH_i\brt+\bH_r\brt  &\quad\quad (r>R_2)\\
        \bH_s\brt  &\quad\quad (R_1<r<R_2)\\
        \bH_t\brt  &\quad\quad (0<r<R_1)
    \end{cases} \quad,
\end{equation}
where the subscript/superscript $r,s,t$ stand for ``$reflected$'', ``$shell$'', and ``$transmitted$'', respectively.

Following Stratton \cite{Stratton}, the incident plane wave can be written as,
\begin{subequations}\label{eq:incident}
  \begin{eqnarray}
      \bE_i\brt &=& \xu E_0 e^{ik_0z-i\omega t} = E_0 e^{-i\omega t}\sum_{n=1}^{\infty} i^n\frac{2n+1}{n(n+1)}\Big[ \mf^{(1)}_{o1n}(k_0\br) - i \nf^{(1)}_{e1n}(k_0\br) \Big]\\
      \bH_i\brt &=& \yu\frac{k_0}{\omega}E_0 e^{ik_0z-i\omega t} = -\frac{k_0E_0}{\omega} e^{-i\omega t}\sum_{n=1}^{\infty} i^n\frac{2n+1}{n(n+1)}\Big[ \mf^{(1)}_{e1n}(k_0\br) + i \nf^{(1)}_{o1n}(k_0\br) \Big],
  \end{eqnarray}
\end{subequations}
where $k_0=2\pi/\lambda=\omega/c$ is the wave vector. The two special functions are defined in spherical coordinates as,
\begin{subequations}\label{eq:mnfunc}
    \begin{eqnarray}
        \mf^{(q)}_{{o \atop e}mn}(k\br) &=& \pm~\thetau~\frac{m}{\sin\theta}z_n^{(q)}(kr)P_n^m(\cos\theta) {\cos\atop\sin}m\phi - \phiu~z_n^{(q)}(kr)\frac{\partial P_n^m(\cos\theta)}{\partial\theta} {\sin\atop\cos}m\phi  \\
        \nf^{(q)}_{{o \atop e}mn}(k\br) &=& \ru~\frac{n(n+1)}{kr}z_n^{(q)}(kr)P_n^m(\cos\theta){\sin\atop\cos}m\phi + \thetau~\frac{1}{kr} \big[kr~z_n^{(q)}(kr)\big]'~\frac{\partial P_n^m(\cos\theta)}{\partial\theta}{\sin\atop\cos}m\phi \nonumber\\
        &\quad& \pm~\phiu~\frac{m}{kr\sin\theta}\big[kr~z_n^{(q)}(kr)\big]'~P_n^m(\cos\theta){\cos\atop\sin}m\phi ~~,
    \end{eqnarray}
\end{subequations}
where the superscripts $q=(1,2,3,4)$ refer to the four spherical Bessel and Hankel functions $\Big(j_n,y_n,h^{(1)}_n,h^{(2)}_n\Big)$ that replace $z_n^{(q)}$, respectively, and the prime denotes differentiation with respect to the argument $kr$. Since the incident field contains only $m=1$ terms, all $m \ne 1$ terms are trivial in our derivation.

As within the original Mie theory \cite{Mie,Stratton}, the field in vacuum, \emph{i.e.}, outside of the core-shell structure (``reflected'' field), can be written as,
\begin{subequations}\label{eq:reflected}
    \begin{eqnarray}
        \bE_r\brt &=&  E_0 e^{-i\omega t}\sum_{n=1}^{\infty} i^n\frac{2n+1}{n(n+1)}\Big[ a_n^r\mf^{(3)}_{o1n}(k_0\br) - i b_n^r \nf^{(3)}_{e1n}(k_0\br) \Big]\\
        \bH_r\brt &=&  -\frac{k_0E_0}{\omega} e^{-i\omega t}\sum_{n=1}^{\infty} i^n\frac{2n+1}{n(n+1)}\Big[ b_n^r \mf^{(3)}_{e1n}(k_0\br) + i a_n^r \nf^{(3)}_{o1n}(k_0\br) \Big],
    \end{eqnarray}
\end{subequations}
with undetermined coefficients $(a_n^r, b_n^r)$. The spherical Hankel functions of the first kind ($q=3$) are chosen for outgoing waves. The field inside the core (``transmitted'' field) can be written as,
\begin{subequations}\label{eq:transmitted}
    \begin{eqnarray}
        \bE_t\brt &=&  E_0 e^{-i\omega t}\sum_{n=1}^{\infty} i^n\frac{2n+1}{n(n+1)}\Big[ a_n^t\mf^{(1)}_{o1n}(k_1\br) - i b_n^t \nf^{(1)}_{e1n}(k_1\br) \Big]\\
        \bH_t\brt &=&  -\frac{k_1E_0}{\omega} e^{-i\omega t}\sum_{n=1}^{\infty} i^n\frac{2n+1}{n(n+1)}\Big[ b_n^t \mf^{(1)}_{e1n}(k_1\br) + i a_n^t \nf^{(1)}_{o1n}(k_1\br) \Big],
    \end{eqnarray}
\end{subequations}
with undetermined coefficients $(a_n^t, b_n^t)$. The spherical Bessel functions ($q=1$) are chosen for convergence at the origin. $k_1 = n_1k_0$ is the wave vector inside the core.

The main modification to the original Mie theory is the field in the spherical shell ($R_1 < r < R_2$). Since the convergence requirements at the origin (for ``transmitted'' field) and infinity (for ``reflected'' field) are absent in the shell, two types of Bessel functions need to be included. Here we choose the spherical Bessel functions ($q=1$, associated with superscript ``$j$'') and spherical Neumann functions ($q=2$, associated with superscript ``$y$''), such that, 
\begin{subequations}\label{eq:shell}
    \begin{eqnarray}
        \bE_s\brt &=&  E_0 e^{-i\omega t}\sum_{n=1}^{\infty} i^n\frac{2n+1}{n(n+1)}\Big[ a_n^j\mf^{(1)}_{o1n}(k_2\br) + a_n^y\mf^{(2)}_{o1n}(k_2\br) - i b_n^j \nf^{(1)}_{e1n}(k_2\br) - i b_n^y \nf^{(2)}_{e1n}(k_2\br) \Big]\\
        \bH_s\brt &=&  -\frac{k_2E_0}{\omega} e^{-i\omega t}\sum_{n=1}^{\infty} i^n\frac{2n+1}{n(n+1)}\Big[ b_n^j \mf^{(1)}_{e1n}(k_2\br) + b_n^y \mf^{(2)}_{e1n}(k_2\br) + i a_n^j \nf^{(1)}_{o1n}(k_2\br) + i a_n^y \nf^{(2)}_{o1n}(k_2\br) \Big],
    \end{eqnarray}
\end{subequations}
with undetermined coefficients $(a_n^j, a_n^y, b_n^j, b_n^y)$. $k_2 = n_2k_0$ is the wave vector inside the shell.

Determining the total field in \re{eq:total} requires solving for the set of coefficients $(a_n^r, a_n^j, a_n^y, a_n^t, b_n^r, b_n^j, b_n^y, n_n^t)$ by applying the boundary conditions at two surfaces,
\begin{subequations}\label{eq:boundary}
    \begin{eqnarray}
        {\ru\times\big[\bE_i\brt+\bE_r\brt\big] = \ru\times\bE_s\brt
        \atop
        \ru\times\big[\bH_i\brt+\bH_r\brt\big] = \ru\times\bH_s\brt} \Bigg|_{r=R_2} \\
        {\ru\times\bE_s\brt = \ru\times\bE_t\brt
        \atop
        \ru\times\bE_s\brt = \ru\times\bH_t\brt} \Bigg|_{r=R_1}  \quad.
    \end{eqnarray}
\end{subequations}
Plugging \res{eq:incident}, (\ref{eq:reflected}), (\ref{eq:shell}), and (\ref{eq:transmitted}) into \res{eq:boundary} and comparing each order of $n$ yields,
\begin{subequations}
    \begin{eqnarray}
        &&\ru\times\big[\mf^{(1)}_{o1n}(k_0\br)+a_n^r\mf^{(3)}_{o1n}(k_0\br) - i\nf^{(1)}_{e1n}(k_0\br) - ib_n^r\nf^{(3)}_{e1n}(k_0\br)\big]  \nonumber\\
        &&\quad\quad\quad\quad\quad\quad= \ru\times\big[a_n^j\mf^{(1)}_{o1n}(k_2\br)+a_n^y\mf^{(2)}_{o1n}(k_2\br) - ib_n^j\nf^{(1)}_{e1n}(k_2\br) - ib_n^y\nf^{(2)}_{e1n}(k_2\br)\big] \quad\Big|_{r=R_2} \\
        &&\ru\times k_0\big[\mf^{(1)}_{e1n}(k_0\br)+b_n^r\mf^{(3)}_{e1n}(k_0\br) + i\nf^{(1)}_{o1n}(k_0\br) + ia_n^r\nf^{(3)}_{o1n}(k_0\br)\big]  \nonumber\\
        &&\quad\quad\quad\quad\quad\quad= \ru\times k_2\big[b_n^j\mf^{(1)}_{e1n}(k_2\br)+b_n^y\mf^{(2)}_{e1n}(k_2\br) + ia_n^j\nf^{(1)}_{o1n}(k_2\br) + ia_n^y\nf^{(2)}_{o1n}(k_2\br)\big] \quad\Big|_{r=R_2} \\
        &&\ru\times\big[a_n^j\mf^{(1)}_{o1n}(k_2\br)+a_n^y\mf^{(2)}_{o1n}(k_2\br) - ib_n^j\nf^{(1)}_{e1n}(k_2\br) - ib_n^y\nf^{(2)}_{e1n}(k_2\br)\big] \nonumber\\
        &&\quad\quad\quad\quad\quad\quad= \ru\times\big[ a_n^t\mf^{(1)}_{o1n}(k_1\br) - i b_n^t \nf^{(1)}_{e1n}(k_1\br) \big] \quad\Big|_{r=R_1}\\
        &&\ru\times k_2\big[b_n^j\mf^{(1)}_{e1n}(k_2\br)+b_n^y\mf^{(2)}_{e1n}(k_2\br) + ia_n^j\nf^{(1)}_{o1n}(k_2\br) + ia_n^y\nf^{(2)}_{o1n}(k_2\br)\big] \nonumber\\
        &&\quad\quad\quad\quad\quad\quad= \ru\times k_1\big[ b_n^t \mf^{(1)}_{e1n}(k_1\br) + i a_n^t \nf^{(1)}_{o1n}(k_1\br) \big] \quad\Big|_{r=R_1}  \quad.
    \end{eqnarray}
\end{subequations}
Since the above equations must hold for all angles $\theta$ and $\phi$, we find eight linear equations for the eight undetermined coefficients $(a_n^r, a_n^j, a_n^y, a_n^t, b_n^r, b_n^j, b_n^y, n_n^t)$,
\begin{subequations}\label{eq:lns}
    \begin{eqnarray}
        j_n(k_0R_2)+a_n^rh_n^{(1)}(k_0R_2) &=& a_n^jj_n(k_2R_2) + a_n^yy_n(k_2R_2) \\
        \frac{\big[k_0R_2~j_n(k_0R_2)\big]'}{k_0R_2}+b_n^r\frac{\big[k_0R_2~h_n^{(1)}(k_0R_2)\big]'}{k_0R_2} &=& b_n^j\frac{\big[k_2R_2~j_n(k_2R_2)\big]'}{k_2R_2} + b_n^y\frac{\big[k_2R_2~y_n(k_2R_2)\big]'}{k_2R_2} \\  
        k_0j_n(k_0R_2)+b_n^rk_0h_n^{(1)}(k_0R_2) &=& b_n^jk_2j_n(k_2R_2) + b_n^yk_2y_n(k_2R_2) \\
        k_0\frac{\big[k_0R_2~j_n(k_0R_2)\big]'}{k_0R_2}+a_n^rk_0\frac{\big[k_0R_2~h_n^{(1)}(k_0R_2)\big]'}{k_0R_2} &=& a_n^jk_2\frac{\big[k_2R_2~j_n(k_2R_2)\big]'}{k_2R_2} + a_n^yk_2\frac{\big[k_2R_2~y_n(k_2R_2)\big]'}{k_2R_2} \\  
        a_n^jj_n(k_2R_1) + a_n^yy_n(k_2R_1) &=&  a_n^tj_n(k_1R_1)\\
        b_n^j\frac{\big[k_2R_1~j_n(k_2R_1)\big]'}{k_2R_1} + b_n^y\frac{\big[k_2R_1~y_n(k_2R_1)\big]'}{k_2R_1} &=& b_n^t\frac{\big[k_1R_1~j_n(k_1R_1)\big]'}{k_1R_1}\\  
        b_n^jk_2j_n(k_2R_1) + b_n^yk_2y_n(k_2R_1) &=& b_n^tk_1j_n(k_1R_1)\\
        a_n^jk_2\frac{\big[k_2R_1~j_n(k_2R_1)\big]'}{k_2R_1} + a_n^yk_2\frac{\big[k_2R_1~y_n(k_2R_1)\big]'}{k_2R_1} &=& a_n^tk_1\frac{\big[k_1R_1~j_n(k_1R_1)\big]'}{k_1R_1}    \quad.
    \end{eqnarray}
\end{subequations}
These equation uniquely specify the undetermined coefficient for each order $n$. Solving \res{eq:lns} for these variables using any linear algebra package thus gives the full solution of the total field in \re{eq:total}.

This method can further be generalized to core-shell structures with more than one layer of shells. For each additional layer, four undetermined coefficients and one additional set of surface-matching conditions need to be included, adding four more linear equations in \res{eq:lns}.

\section{Supplementary Note 4: Discussion of the field enhancement uncertainty caused by the manufacturing size dispersity}

In \rf{fig:3}a and \rf{fig:4}d-f of the main text, we demonstrated the field enhancement for core-shell structures, with an uncertainty (red shaded area) caused by the manufacturing size dispersity. To be specific, the core-shell structures used in the experiment have an outer diameter of $D_2 = 147 \pm 7$ nm and an inner diameter of $D_1 = 118 \pm 4$ nm. This dispersity causes a rather large variation in the field enhancement. Here, we use \rf{fig:4}d as an example to illustrate how this variation is addressed.

In \rf{fig:SI2}, the dashed line is the field enhancement $|\alpha|^2$ simulated for $D_2 = 147 - 7$ nm and $D_1 = 118 + 4$ nm, where the shell thickness is the thinnest within the manufacturing dispersity. In this case, the peak field enhancement ($|\alpha|^2\sim110$) is obtained at approximately 700 nm. On the other extreme, for $D_2 = 147 + 7$ nm and $D_1 = 118 - 4$ nm, where the shell is the thickest, the peak field enhancement is significantly lower ($\sim$ 40), and the position is blue-shifted to approximately 610 nm. It is clear that the field-enhancement profile is the closest to that of the solid Au nanospheres of the same size when the shell thickness is the thickest possible within the manufacturing dispersity. 

Since the small manufacturing dispersity causes pronounced differences in the field enhancement (almost 3 times in peak height and a blue shift of $\approx$ 90 nm), it is inaccurate to show just {\em one} line of field enhancement for a particular combination of outer and inner radii. Instead, at each wavelength, we calculated the range of field enhancement for all possible combinations of outer and inner radii within the manufacturing dispersity (shown as the red shaded area in \rf{fig:SI2}) and highlight the average value (shown as the red circled line). In contrast to core-shell structures, our simulation shows that the same uncertainty for solid Au nanospheres is negligible.

\subsection{Supplementary References}
\bibliographystyle{apsrev4-2}
\bibliography{NL}